\documentclass[11pt,aps,prl,twocolumn,superscriptaddress,groupedaddress]{revtex4}
\usepackage[utf8]{inputenc}
\usepackage{amsmath}
\usepackage{amsfonts}
\usepackage{amssymb}
\usepackage{hyperref}
\usepackage{graphicx}
\usepackage{setspace}

\begin{document}

\title{An observation of circular RNAs in bacterial RNA-seq data.}

\author{Nicolas Innocenti}\affiliation{Department of Computational Biology, KTH Royal Institute of Technology, AlbaNova University Center, Roslagstullsbacken 17, SE-10691 Stockholm, Sweden}\affiliation{Combient AB, Nettovägen 6, SE-17541, Järfälla, Sweden}
\author{ Hoang-Son Nguyen}\affiliation{System Biology Programme, KTH Royal Institute of Technology, SE-100 44 Stockholm, Sweden}
\author{Aymeric Fouquier~d'H\'erou\"el}\affiliation{Luxembourg Centre for Systems Biomedicine, University of Luxembourg, 7,~Avenue des Hauts-Fourneaux, L-4362 Esch-sur-Alzette, Luxembourg}
\author{Erik Aurell}\affiliation{Department of Computational Biology, KTH Royal Institute of Technology, AlbaNova University Center, Roslagstullsbacken 17, SE-10691 Stockholm, Sweden}\affiliation{Department of Information and Computer Science, Aalto University, Konemiehentie 2, FI-02150 Espoo, Finland}
\singlespace

\begin{abstract}
Circular RNAs (circRNAs) are a class of RNA with an important role in micro RNA (miRNA) regulation recently discovered in Human and various other eukaryotes as well as in archaea. Here, we have analyzed RNA-seq data obtained from {\it Enterococcus faecalis} and {\it Escherichia coli} in a way similar to previous studies performed on eukaryotes. We report observations of circRNAs in RNA-seq data that are reproducible across multiple experiments performed with different protocols or growth conditions.
\end{abstract}
\maketitle
\singlespace

Circular RNAs (circRNAs) are a type of RNA transcripts the 3'ends of which are ligated to their 5'ends during splicing. They were first discovered in plants \cite{Sanger01111976} almost 40 years ago and relatively little studied for several decades. Recently, an incidental discovery of a high abundance of circRNAs in human \cite{10.1371/journal.pone.0030733} and their role in micro RNA (miRNA) regulation \cite{Hansen:2013fj} attracted the attention of many. Large scale searches for novel circRNAs was performed in animals \cite{Memczak:2013sf} and archaea \cite{Danan02122011} leading to hundreds of previously unknown circularised transcripts.  While it has been shown that splicing in bacteria occurs, though much more rarely than in eukaryotes \cite{Woodson01051998}, that circRNAs can be synthesized by bacteria in vivo using genetic engineering, and that those transcripts can even direct translations of proteins in those organisms, it is commonly believed that all natural transcripts in bacteria are linear \cite{Perriman01091998}.  

Our starting point is a slightly modified version of the well established pipeline used by Memczak et al. \cite{Memczak:2013sf} to analyse our recently published transcriptome data on \textit{E. faecalis} v583 \cite{Innocenti2014}. The major difference of our version of the pipeline is that we removed the requirement for break points to be flanked by GU/AG sites, as such a configuration is characteristic of spliceosomal splicing, absent in bacteria. As a result, the algorithm reports as potential circRNA all cases where the 3' region of a read can be mapped upstream of its 5' region and where the sequence on the reference genome near them corresponds to the one in the read insert (Figure 1(a) ).


The discovery pipeline was applied to the transcriptome coined "IlluminaSt" [SRA accession ID : SRR1770393] in our previous nomenclature \cite{Innocenti2014}, leading to a total of 134 circRNA candidates (Table S1). Based on these, we built a custom database of junctions encompassing 250 nt on each side of the predicted junctions (Figure 1(a)). This database was subsequently used as reference against which we aligned whole reads directly. After alignment, we counted the number of reads that spanned the junction, and reads extending 8 nt or more on both sides of a junction were termed "long spanning reads" (Figure 1(b)). 

\begin{figure*}[ht]
\centering
\centering
\includegraphics[width=.9\textwidth]{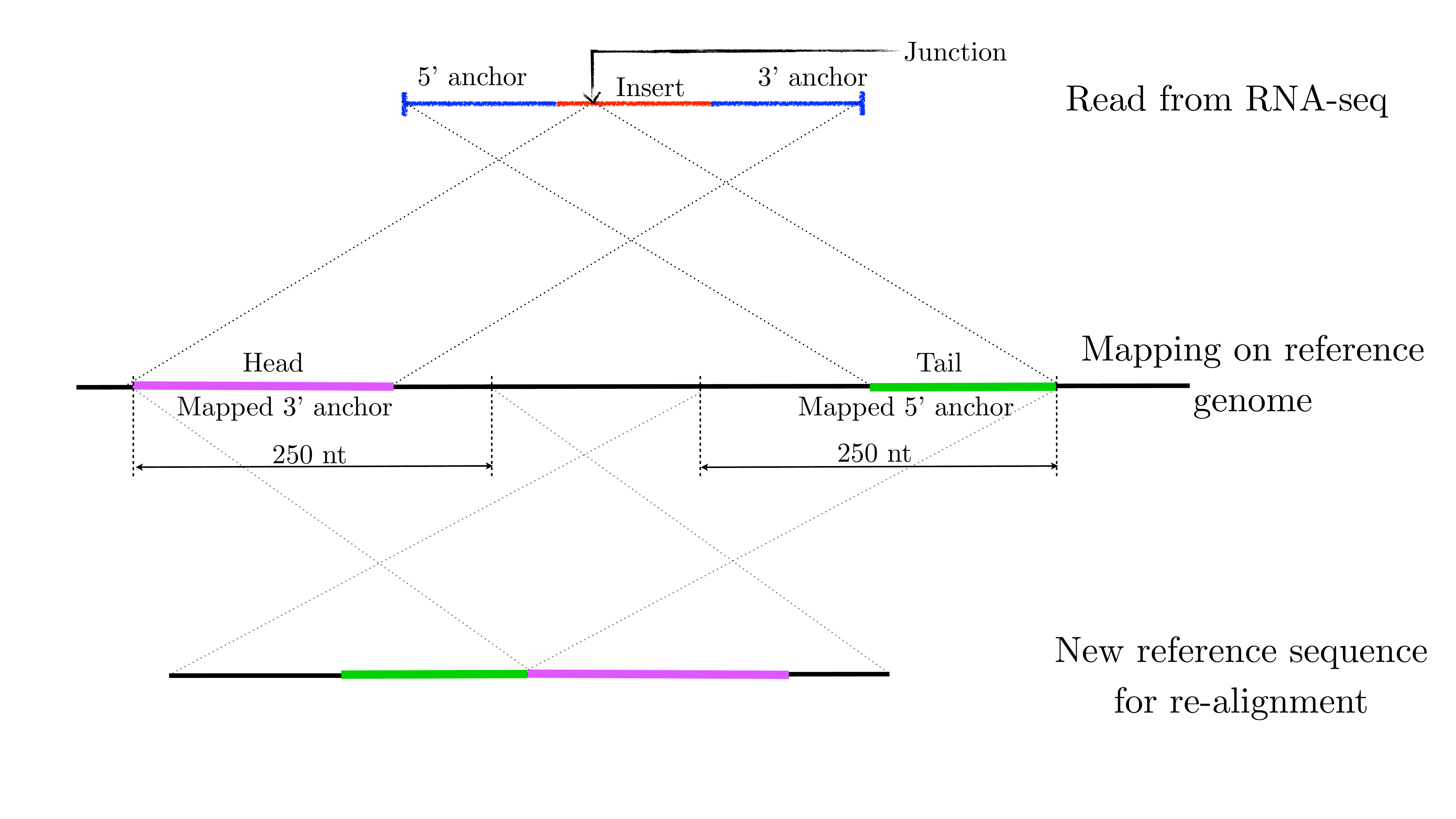}\ \\ 
(a)\\
\includegraphics[width=.9\textwidth]{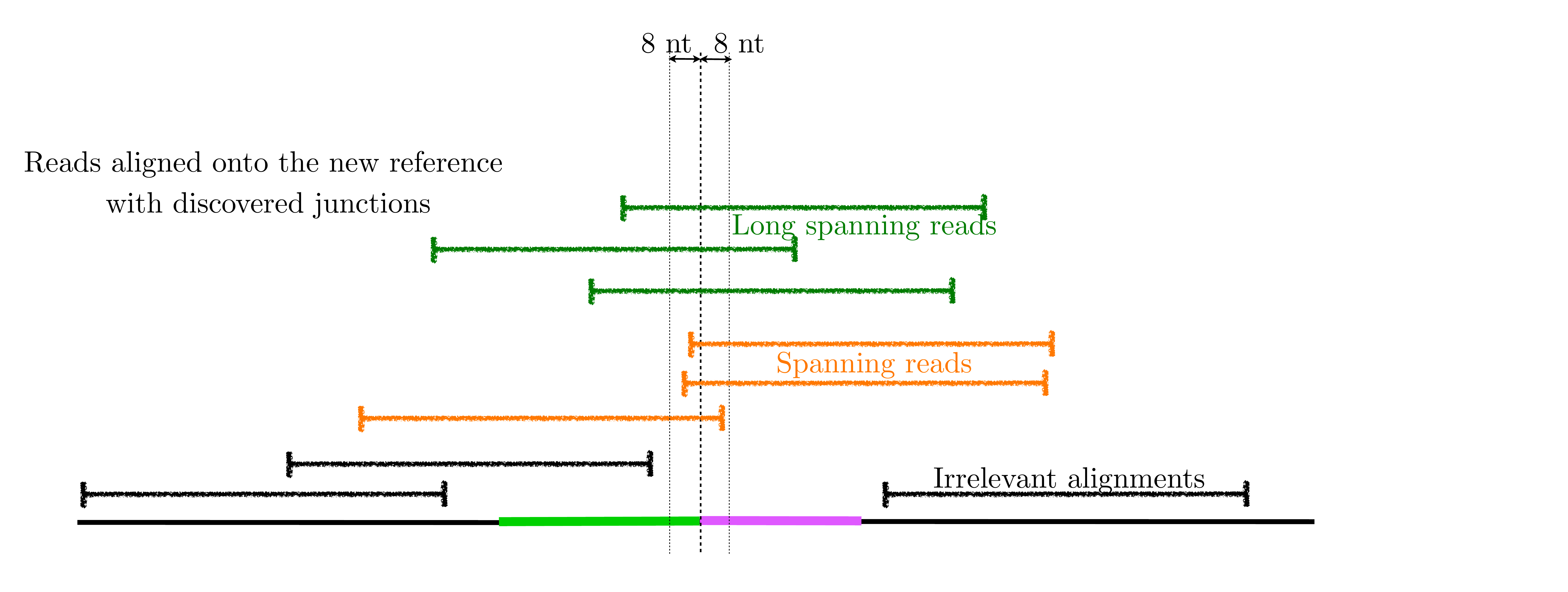}\ \\ 
(b)\\
\caption{(a) In the discovery phase, reads are cut in silico into three parts : a region of 20 nt at the 5'end (5'anchor), another of 20 nt at the 3'end (3'anchor) and the middle region containing the remaining nucleotides of the reads. For reads that contain a tail-to-head junction the insert, the discovery pipeline will map the 5'anchor downstream of the 3'anchor and then use the insert sequence to recover the beginning and end of the RNA transcript before circularisation. Once this position has been found, a new reference sequence is created to be used for a second mapping procedure. (b) Schematic representation of reads after aligning to the new reference built during the discovery phase. Reads that map entirely on one side of the junction are irrelevant to our purpose. Each read that maps across the junction votes in favor of its existence, with a higher confidence for reads having a longer sequences on each side. }
\end{figure*}

Alignment of the same "IlluminaSt" transcriptome reads yielded an enrichment of long spanning reads  (Table S1) caused by relaxed constraints since anchors in our 50 nt long reads were required to be 20 nt long: a junction needed to be in a region of 10 nt to be identified by the detection algorithm, which, assuming uniform RNA fragmentation, would happen in 20\% of the cases. 
Few large deviations from the typically observed two-to-ten-fold enrichment (Table S1) are plausibly explained by fragmentation biases in the sequencing sample preparation \cite{Roberts2011}. 

In addition, we mapped reads from four other \emph{E. faecalis} V583 transcriptomes \cite{Innocenti2014} named "KTH", "KTHr", "Rt" and "St"  [SRA accession IDs: SRR1766416, SRR1769261, SRR1769749, SRR1769750] acquired using different experimental protocols, growth conditions and sequencing platforms. Out of the previously mentioned 134 candidates, 81 (60\%) were confirmed by at least two long spanning reads from at least two transcriptomes (Table S1).

Removing cases with multiply mapped read fragments and retaining only candidates supported by at least five non-duplicated reads, we extracted a list of 23 transcripts with lengths between 56 and 549 nt for further analysis (Table S1). Among these, 15 (65\%) were found with long spanning reads in at least two transcriptomes, and only one case (circ\_000082 corresponding to the sRNA ref114) was observed only in "IlluminaSt". We then verified that, while antisense transcription was observed in the neighborhood of several candidates, in all but one case no long antisense spanning reads were observed, with the exception of the candidate "circ\_000104", discussed in more detail below. We retained circ\_000104 because it was found in all transcriptomes analyzed, had the highest number of long-spanning reads in the KTH transcriptome, and the antisense signal across its junction was about 20 times weaker than the direct signal. In all the 23 cases, coverage signal clearly indicates presence of transcripts across the entire candidate sequences (See "The ppRNomebrowse  "\footnote{\url{http://ebio.u-psud.fr/eBIO_BDD.php}}\cite{Innocenti2014}).

In order to estimate the false discovery rate of our procedure, we generated a database of 5000 tail-to-head junctions up to 600 nt apart randomly distributed over the genome with a uniform distribution and assessed the number of long spanning reads found covering those junctions. Indeed not a single one was reported while over 95\% the database had reads aligned in an irrelevant way near the junction, strengthening the confidence in our analysis of the true transcriptomic data.


Of the 23 retrieved circRNA candidates, 11 correspond to short RNAs (sRNAs) previously described, the majority of which having so far uncharacterised functions, 5 overlap with untranslated regions (UTRs) sometimes including part of a neighbouring coding region, 6 roughly match the locations of annotated genes (including two candidates in the vicinity of {\it EF\_0104/arcA}) and 1 coincides with a tryptophan transfer RNA ({\it tRNA-Trp1}) (TableS1).

\begin{figure*}[ht]
\centering
\includegraphics[width=\textwidth]{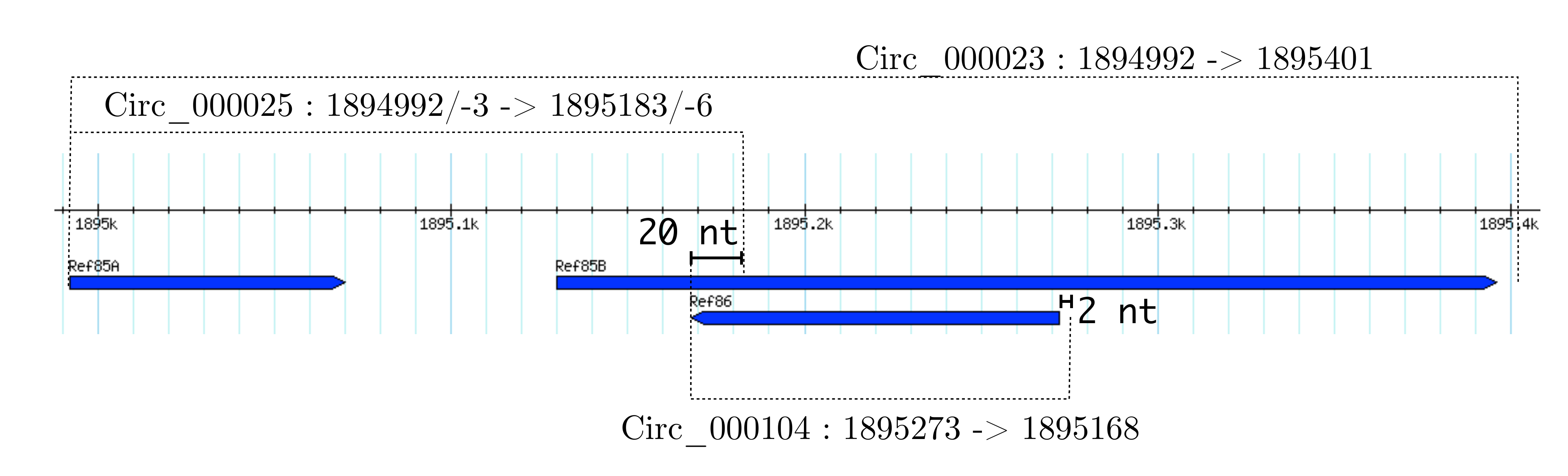}
\caption{Three head-to-tail junctions detected in a region of the genome of {\it E. faecalis} containing previously reported ncRNAs region, suggesting an intricate interaction between the RNA transcripts.\label{fig:Ref8586}}
\end{figure*}

A striking observation is a set of 3 candidates (circ\_000023, circ\_000025 and circ\_000104) in a region we recently reported as containing an antisense organisation of two or three ncRNAs called Ref85A/B and Ref86 \cite{Innocenti2014}. On the forward strand, we observes two tail-to-head junctions linking coordinates 1.894.992 to 1.895.401 and 1.894.992/-3 to 1.895.183/-6, respectively. Position 1.894.992 coincides with the mapped transcription start site (TSS) of Ref85A\cite{Innocenti2014}. The other coordinates correspond to the 3' of Ref85B and to a position 20 nt upstream of the 3' of Ref86 on the opposite strand (Figure \ref{fig:Ref8586}). On the reverse strand, the detected junction links position 1.895.168 to 1.895.273, i.e., 2 nt upstream of the reported beginning of Ref86 to exactly its reported end.

While the role of such an intricate organization remains elusive, the above observation clarifies that Ref85A and Ref85B originate from the same primary transcript and that Ref86 may indeed interact with Ref85A/B. 

While it is well known that artifacts in the reverse transcription, particularly template switching, may lead to junction reads similar to those resulting from splicing or circularisation, those effects are expected to be random \cite{Jeck2014}. The presence of non-identical (long) spanning reads across multiple experiments is an argument against  artifactual origin. 
Moreover, the clear tendency for the candidates to be found in non-protein coding regions (UTRs and ncRNAs) that constitute a minor portion of the genome strengthens the biological origin of our observations.

To verify that the effect is not limited to the particular bacterial strain used or to our experimental protocol, we then applied our discovery pipeline to 23 publicly available RNA-seq datasets of {\it E. coli}. Datasets were taken from the SRA database and chosen arbitrarily among experiments done with single-end strand-specific standard RNA-seq on the Illumina sequencing platform, on strain MG1655 or closely related ones, and preferentially performed at high coverage (accession numbers for the datasets and results of the analyses are given in table S2).

From the 23 transcriptomes, we obtained a total of about 20 000 predictions that cluster in 2660 candidate loci (Table S2), indicating that predictions accross different samples lead to similar positions. Among those 2660 loci, 316 were observed in at least 7 transcriptomes (Table S2).

To provide a global overview, we plotted the density of predictions across all 23 samples in Figure 2 and, in Figure S1, the coverage signal calculated from the reads on which predictions are based for the 316 regions.  
The two strongest peaks in the plots correspond to {\it csrB} and {\it csrC}, two short non-coding RNAs with the same function : binding multiple copies of the CsrA protein, which is itself an inhibitor of translation,  in order to inhibit its function and consequently increase protein synthesis. Both have similar structures, forming many hair-pins on an otherwise opened backbone \cite{Gardner01012009}. Smaller peaks in Figure 2 correspond to {\it hdeA} and {\it B}, an operon coding for two well known proteins in {\it E. coli}, {\it yehD}, a region annotated as putative protein, and rnpB, the catalytic subunit of RNase P. The latter is also found circularised in {\it E. faecalis} as candidates circ\_000068, and especially circ\_000048 which is one the strongest signals discussed above.



To summarize, we discovered a series of potential circular RNA candidates in bacteria using publicly available discovery pipeline and RNA-seq data. We have verified that a number of the candidates appear to be reproducible across experiments with different RNA-seq protocols and platforms and the case of two or possibly three ncRNAs antisense in {\it E. faecalis} to each other showing signs of circularisation in an intricate manner.  

A result of our analysis is that most candidates were not found in coding regions but rather in functional RNA transcripts or UTRs. In many cases, candidates found were reproducible across experiments with good accuracy. In particular the rnpB sRNA shows one of the strongest signals in both {\it E. faecalis} and {\it E. coli}. We finally remark that in all samples, the signal from putative circRNA was weak compared to the total amount of RNA detected in the corresponding regions, suggesting that either the circularized form is rare byproduct of a post-transcriptional process or that the circular form is not well captured by the RNA-seq procedure, for instance by escaping fragmentation.

\begin{figure*}[htb]
\centering
\includegraphics[width=.9\textwidth]{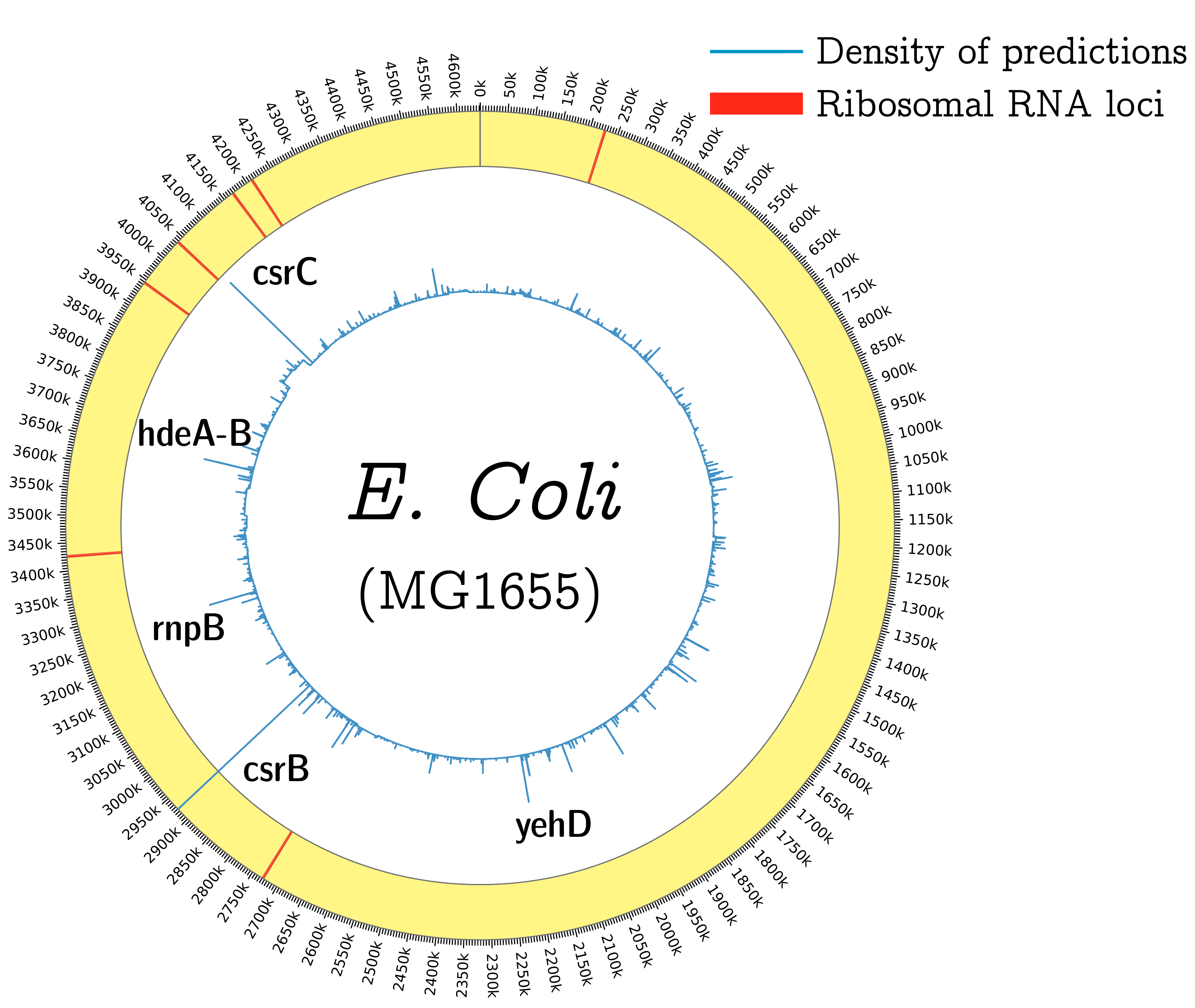}
\caption{Density of predicted circular RNAs in E. coli for the 23 {\it E. coli} transcriptomes analysed. The plot uses a linear scale and is normalised to the maximum value across the genome (the peak at {\it csrB}). \label{fig:Coli}}
\end{figure*}
 
\section*{Authors' Contributions}
A.F.d’H. and N.I. formulated the research problem. N.I., A.F.d’H., and E.A. designed the work. H.-S.N. implemented the computational pipeline. NI and H.-S.N. performed data analysis. N.I. and E.A. wrote the paper. All authors critically reviewed the manuscript.


\clearpage

\begin{figure*}[htb]
\centering
\includegraphics[height=.9\textwidth,angle =90]{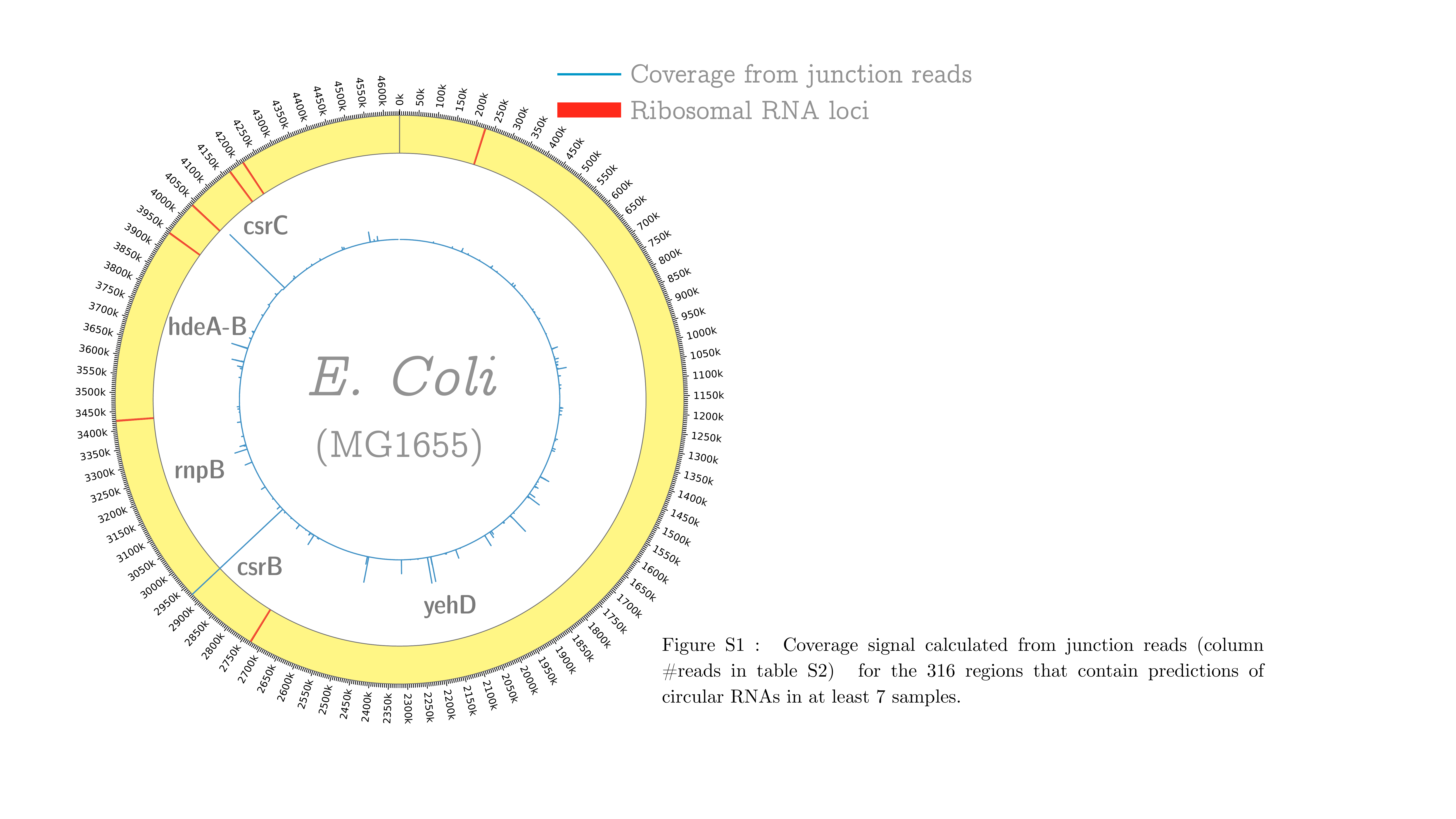}
\end{figure*}

\end{document}